\documentstyle[12pt,epsf]{article}
\def\dsppp{$D^+_s\rightarrow \pi^+\pi^-\pi^+$}
\def\dsrhopi{$D^+_s\rightarrow \rho^0\pi^+$}
\def\dsfpi{$D^+_s\rightarrow f^{}_0(980)\pi^+$}
\def\dsfheavypi{$D^+_s\rightarrow f^{}_0(1370)\pi^+$} 
\begin{document}
\Large
\begin{center}
 Study of the $D^+_s \to \pi^- \pi^+ \pi^+$ decay and measurement of $f_0$
 masses and widths.\end{center}
\normalsize
\small\begin{center}
    E.~M.~Aitala,$^9$
       S.~Amato,$^1$
    J.~C.~Anjos,$^1$
    J.~A.~Appel,$^5$
       D.~Ashery,$^{14}$
       S.~Banerjee,$^5$
       I.~Bediaga,$^1$
       G.~Blaylock,$^8$
    S.~B.~Bracker,$^{15}$
    P.~R.~Burchat,$^{13}$
    R.~A.~Burnstein,$^6$
       T.~Carter,$^5$
    H.~S.~Carvalho,$^{1}$
    N.~K.~Copty,$^{12}$
    L.~M.~Cremaldi,$^9$
       C.~Darling,$^{18}$
       K.~Denisenko,$^5$
       S.~Devmal,$^3$
       A.~Fernandez,$^{11}$
    G.~F.~Fox,$^{12}$
       P.~Gagnon,$^2$
       C.~Gobel,$^1$
       K.~Gounder,$^9$
    A.~M.~Halling,$^5$
       G.~Herrera,$^4$
       G.~Hurvits,$^{14}$
       C.~James,$^5$
    P.~A.~Kasper,$^6$
       S.~Kwan,$^5$
    D.~C.~Langs,$^{12}$
       J.~Leslie,$^2$
       B.~Lundberg,$^5$
       J.~Magnin,$^1$       
       A.~Massafferri,$^1$
       S.~MayTal-Beck,$^{14}$
       B.~Meadows,$^3$
 J.~R.~T.~de~Mello~Neto,$^1$
       D.~Mihalcea,$^7$
    R.~H.~Milburn,$^{16}$
    J.~M.~de~Miranda,$^1$
       A.~Napier,$^{16}$
       A.~Nguyen,$^7$
    A.~B.~d'Oliveira,$^{3,11}$
       K.~O'Shaughnessy,$^2$
    K.~C.~Peng,$^6$
    L.~P.~Perera,$^3$
    M.~V.~Purohit,$^{12}$
       B.~Quinn,$^9$
       S.~Radeztsky,$^{17}$
       A.~Rafatian,$^9$
    N.~W.~Reay,$^7$
    J.~J.~Reidy,$^9$
    A.~C.~dos Reis,$^1$
    H.~A.~Rubin,$^6$
    D.~A.~Sanders,$^9$
 A.~K.~S.~Santha,$^3$
 A.~F.~S.~Santoro,$^1$
       A.~J.~Schwartz,$^{3}$
       M.~Sheaff,$^{17}$
    R.~A.~Sidwell,$^7$
    A.~J.~Slaughter,$^{18}$
    M.~D.~Sokoloff,$^3$
       J.~Solano,$^1$
    N.~R.~Stanton,$^7$
    R.~J.~Stefanski,$^5$  
       K.~Stenson,$^{17}$ 
    D.~J.~Summers,$^9$
       S.~Takach,$^{18}$
       K.~Thorne,$^5$
    A.~K.~Tripathi,$^{7}$
       S.~Watanabe,$^{17}$
 R.~Weiss-Babai,$^{14}$
       J.~Wiener,$^{10}$
       N.~Witchey,$^7$
       E.~Wolin,$^{18}$
    S.~M.~Yang,$^7$
       D.~Yi,$^9$
       S.~Yoshida,$^7$
       R.~Zaliznyak,$^{13}$ and
       C.~Zhang$^7$ \\
\normalsize  (Fermilab E791 Collaboration)\\
\small
{\it 
$^1$ Centro Brasileiro de Pesquisas F{\'\i}sicas, Rio de Janeiro, Brazil,
$^2$ University of California, Santa Cruz, California 95064,
$^3$ University of Cincinnati, Cincinnati, Ohio 45221,
$^4$ CINVESTAV, Mexico City, Mexico,
$^5$ Fermilab, Batavia, Illinois 60510,
$^6$ Illinois Institute of Technology, Chicago, Illinois 60616,
$^7$ Kansas State University, Manhattan, Kansas 66506,
$^8$ University of Massachusetts, Amherst, Massachusetts 01003,
$^9$ University of Mississippi-Oxford, University, Mississippi 38677,
$^{10}$ Princeton University, Princeton, New Jersey 08544,
$^{11}$ Universidad Autonoma de Puebla, Puebla, Mexico,
$^{12}$ University of South Carolina, Columbia, South Carolina 29208,
$^{13}$ Stanford University, Stanford, California 94305,
$^{14}$ Tel Aviv University, Tel Aviv, Israel,
$^{15}$ Box 1290, Enderby, British Columbia, V0E 1V0, Canada,
$^{16}$ Tufts University, Medford, Massachusetts 02155,
$^{17}$ University of Wisconsin, Madison, Wisconsin 53706,
$^{18}$ Yale University, New Haven, Connecticut 06511
}\end{center}
\normalsize

\begin{center}{August, 2000}\end{center}

\begin{abstract}
From a sample of 848  $\pm$  44 $D_s^+ \to \pi^- \pi^+ \pi^+$ decays, we find 
 $\Gamma(D_s^+ \to \pi^- \pi^+ \pi^+) / \Gamma(D_s^+ \to \phi \pi^+)  = 
 0.245 \pm 0.028^{+0.019}_{-0.012} $. Using a Dalitz plot analysis 
 of this three body decay, we find   significant contributions from the
  channels $\rho^0(770)\pi^+$, $\rho^0(1450)\pi^+$, $f_0(980)\pi^+$,
 $f_2(1270)\pi^+$,  and  $f_0(1370)\pi^+$. We present also the  values 
 obtained for masses and widths of the resonances $f_0(980)$ and $f_0(1370)$.
\end{abstract}

The charm meson decay \dsppp\ and its charge conjugate 
(implicit throughout this paper) is Cabibbo-favored but 
has no strange meson in the final state. The decay can proceed 
via spectator amplitudes, producing intermediate resonant 
states with hidden strangeness, e.g. \dsfpi, with 
$s\bar{s}$ quarks in the $f_0(980)$.  Also, the decay can
proceed via $W$-annihilation amplitudes,  producing 
intermediate states with no strangeness, e.g. \dsrhopi. A 
$W$-annihilation amplitude could also produce the intermediate 
state \dsfheavypi, assuming the $f_0(1370)$ consists mostly 
of $u\bar{u}$ and $d\bar{d}$ quarks as predicted by the simple 
quark model\cite{pdg}.  To determine the relative importance of these 
different decay mechanisms, one can use an amplitude analysis. 
Such an analysis is also able to determine the masses and decay 
widths of the intermediate states.

In general, scalar resonances have large decay fractions in the 3-body decays 
of $ D $-mesons, and such decays provide a relatively clean laboratory
 in which to study the properties of the scalars. In particular,
 isoscalar intermediate states are dominant in    
$ D_s^+ \to \pi^- \pi^+ \pi^+ $ decays\cite{e691,e687}. The largest  
contribution to this final state comes from the decay involving  
the scalar meson $f_0(980)$ whose nature  is a long-standing puzzle. 
It has been described as a $q \bar q$ state , a $K \bar K$ molecule,
 a glueball, and  a 4-quark state\cite{mont}. 

In this paper we extend the reach of previous studies using
the larger data sample from Fermilab experiment E791.
We present an amplitude analysis which includes a greater number of
possible resonant states, and we measure the masses and widths of
the scalar resonances $ f_0(980) $ and $ f_0(1370) $ with
better precision. Taken together with the results of the companion 
analysis\cite{e791dp}, these results provide new insights into the
characteristics of scalar mesons and their
importance in  charm meson decay.

The data were produced by 500 GeV/$c$ $ \pi^- $ interactions in five thin
foils (one platinum, four diamond) separated by gaps of 1.34 to 1.39~cm. 
The detector, the data set, the reconstruction, and the resulting
vertex resolutions have been described previously\cite{OLDTPL}.
After reconstruction, events with evidence of well-separated
production (primary) and decay  (secondary) vertices were retained for 
further analysis. From the 3-prong secondary vertex candidates,
we select a  $\pi^- \pi^+ \pi^+$ sample with invariant mass ranging 
from 1.7 to 2.1 GeV/c$^2$. For this analysis all charged particles are taken
to be pions; i.e., no direct use is made of particle identification.

\begin{figure}[hbt]
\centerline{\epsfysize=3.00in \epsffile{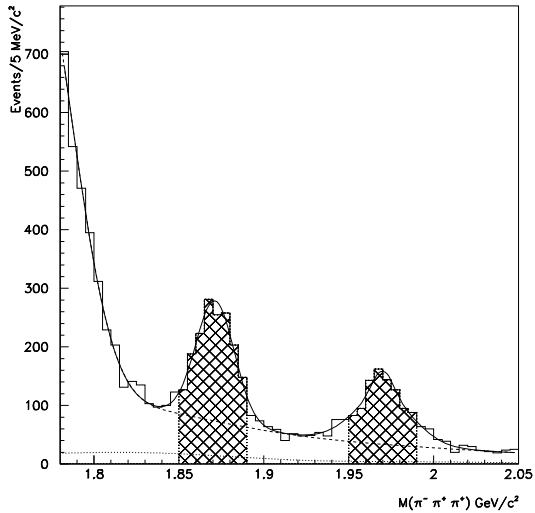}}
\caption{The $\pi^- \pi^+ \pi^+$ effective mass spectrum. The dotted line 
represents the $D^0 \to K^-\pi^+$ plus $ D_s^+ \to \eta^{\prime} \pi^+ $ and the
dashed line is the total backgound. Events in the hatched areas
at the $D_s$ mass are  used for the $D_s$ Dalitz plot analysis in this 
Letter. The hatched area
at the $D^+$ mass is  used for the  analysis in the following companion paper [4].}

\label{fig1}
\end{figure}

We require  a candidate's secondary vertex position
to be cleanly separated from the event's primary vertex position
and from the closest target material. The sum of the momentum vectors of the 
three tracks from this secondary vertex must point to the primary vertex.
The candidate's daughter tracks must pass closer to the secondary vertex than 
to the primary vertex, and must not point back to the primary vertex. 
The resulting invariant mass spectrum is shown in Figure \ref{fig1}.

We fit the spectrum of Figure \ref{fig1} as the sum of $ D^+ $ and $ D_s^+ $ 
signals plus background. To account for the signal's
non Gaussian tails, we model each signal as
the sum of two Gaussian distributions with the same centroidbut different
widths. 
We model the background as the sum of four components: a general
combinatorial background, the reflection of the $D^+ \to K^-\pi^+\pi^+$ decay,
reflections of $D^0 \to K^-\pi^+$ plus one extra track (mostly from the primary
vertex),
and $D_s^+ \to \eta' \pi^+$  followed by
$\eta ' \to \rho^0(770) \gamma$, $\rho^0(770) \to \pi^+\pi^-$. 
The $ D^+ \to K^- \pi^+ \pi^+ $ reflection is located below
1.85 GeV/$c^2$ in the $ \pi^- \pi^+ \pi^+ $ spectrum.
The other charm backgrounds populate the whole $\pi^-\pi^+\pi^+$ spectrum.
We use Monte Carlo (MC) simulations to determine the shape 
of each  identified  charm 
background in the $\pi^- \pi^+ \pi^+$ spectrum.
We assume that the combinatorial background falls exponentially with mass.
The levels of $ D^0 \to K^- \pi^+ $ and $ D_s^+ \to \eta^{\prime} \pi^+ $
backgrounds are determined using charm signal rates measured
in our total event sample and branching ratios taken from the
compilation by the Particle Data Group\cite{pdg}.
The parameters describing the combinatorial background 
and the level of the $D^+ \to K^- \pi^+ \pi^+ $ reflection 
are determined from fitting the $ \pi^- \pi^+ \pi^+ $ distribution.
The mass (centroid) and both Gaussian widths for each signal
float in our fit. The fit  finds 1172  $\pm$  61  $D^+$ events and 
848 $\pm$  44 $D_s^+$ events. 
The $D_s^+$ signal is used for the measurement of the decay rate
 $\Gamma( D_s^+ \to \pi^- \pi^+ \pi^+) $
relative to $\Gamma( D_s^+ \to \phi \pi^+ )$ and for an amplitude
analysis of the $ D_s^+ $ Dalitz plot.

 To minimize systematic effects when calculating the
ratio of efficiencies,  we 
select the $D_s^+ \to \phi\pi^+$, $\phi \to K^-K^+$ signal using the same 
track and vertex quality criteria as for the $D_s^+ \to \pi^- \pi^+ \pi^+$ decay. 
The number of normalization events is found to be 1038  $\pm$  44.

We use MC simulations to correct
the signals for geometrical acceptance and detector efficiency. 
We measure a ratio of efficiencies  $\varepsilon(D^+_s \to \pi^- \pi^+ \pi^+)/
\varepsilon(D_s^+ \to \phi\pi^+) = 1.64\pm0.15$ with values $\varepsilon(D^+_s
\to \pi^- \pi^+ \pi^+)$ and  $\varepsilon(D_s^+ \to \phi\pi^+)$ of about 
2\% and 1\%, respectively.
The error systematic is dominated by uncertainties in the MC model
of $ D$  production and relative efficiencies of our selection criteria.

The branching fraction for  $D_s^+ \to \pi^- \pi^+ \pi^+$ relative to
        that for $D_s^+ \to \phi \pi^+$ is measured to be:

\begin{equation}
{ B(D_s^+ \to \pi^- \pi^+ \pi^+)  \over B(D_s^+ \to \phi \pi^+) } = 
 0.245 \pm 0.028^{+0.019}_{-0.012}\ .   
\end{equation}
\noindent
The first error is statistical. The second is systematic, and
is dominated by uncertainties related to the signal and background shapes 
used in the fit, the background levels, and the sample selection criteria. This
value is  smaller than the other experimental results:
E691 \cite{e691} found $0.44\pm 0.10\pm 0.04$, WA82 \cite{wa82} quoted
$0.33\pm 0.10\pm 0.04$ and E687 \cite{e687} presented $0.33\pm 0.058\pm0.058$. 
The PDG \cite{pdg} presents a value of $0.28\pm 0.06$ from a constrained fit.

\begin{figure}[hbt]
\centerline{\epsfysize=3.00in \epsffile{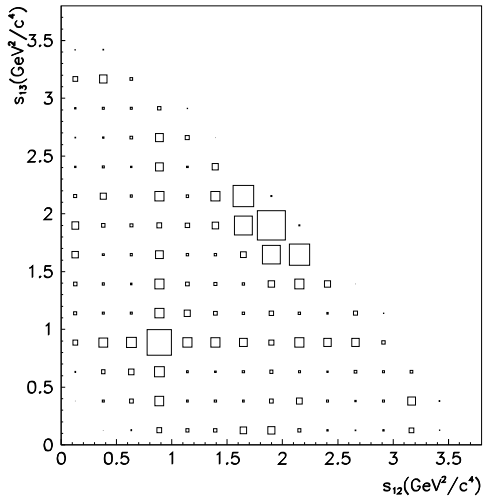}}
\caption{The $D_s^+ \to \pi^- \pi^+ \pi^+$ Dalitz plot. Since there are two
identical particles, the plot is symmetrized.}
\label{fig2}
\end{figure}

 The symmetrized Dalitz plot of the 937 candidates with invariant mass
between 1.95 and 1.99 GeV/c$^2$ is shown in Fig. \ref{fig2}. The integrated signal to background ratio is $ \approx 2 $.
The narrow horizontal and vertical bands of $s_{12}\equiv m^2(\pi^-_1 \pi^+_2)$ 
and $s_{13}\equiv m^2(\pi^-_1 \pi^+_3)$ just below 1 GeV$^2/c^4$
correspond to the $f_0(980) \pi^+$ state with constructive quantum-mechanical 
interference evident where the two bands overlap (the event 
count is four times that of the  individual bands).
At the upper edge of the diagonal, there is another concentration of events
centered at $s_{12} \simeq  s_{13} \simeq $1.8 GeV$^2/c^4$, corresponding to the
$f_2(1270) \pi^+$, $f_0(1370) \pi^+$, and $\rho^0(1450) \pi^+$.

We fit the distribution shown in Figure \ref{fig2} to a signal 
probability distribution function (PDF), which
is a coherent sum of  amplitudes corresponding to the non-resonant decay 
plus five different resonant channels, and a background 
PDF of known shape and magnitude. The resonant channels we include
in the fit are $\rho^0(770) \pi^+$, $f_0(980) \pi^+$, $f_2(1270) \pi^+$, 
$f_0(1370) \pi^+$, and   $\rho^0(1450) \pi^+$. We weight
the signal PDF by the 
acceptance across the Dalitz plot and use the measured line shape and background 
to  determine the ratio of expected signal to background for each event in the 
Dalitz plot.

We assume the non-resonant amplitude to be uniform across the Dalitz plot.
Each resonant amplitude, except that  for the $f_0(980)$, is
parameterized as a product of form factors, a relativistic Breit-Wigner 
function, and an angular momentum amplitude which depends on the spin of the resonance, 

\begin{equation}
{\cal A}_n = \frac {F_D\  ^{J}F_n} {m_{\pi\pi}^2 - m_0^2 + im_0\Gamma(m_{\pi\pi})}
{\cal M}_n^J\ ,
\end{equation}
with
\begin{equation}
\Gamma(m_{\pi\pi}) = \Gamma_0 \frac{m_0}{m_{\pi\pi}} \left(\frac{p^*}{p^*_0}
\right)^{2J+1} \frac{^{J}F_n^2(p^*)}{^{J}F_n^2(p^*_0)}\ .
\end{equation}

The form factors $F_D$ and  $^{J}F_n$ are the Blatt-Weisskopf damping factors
\cite{blatt} respectively for the $D$ and the resonance decays,
$p^*$ is the pion momentum in the resonance rest frame at mass $m_{\pi \pi}\
(p^*_0=p^*(m_0))$ and $J$ is the spin of resonance $n$. ${\cal M}_n^J$ describes
the angular distribution due to the spin\cite{e791dp}. 
Since we have identical particles in the final state, each signal amplitude is 
Bose-symmetrized,  
${\cal A}_n =  {\cal A}_n[({\bf 12}){\bf3}] + {\cal A}_n[({\bf 13}){\bf 2}]$.

For the $f_0(980) \pi^+$ we use a coupled-channel Breit-Wigner function, 
following the parameterization of the WA76 Collaboration\cite{wa76},

\begin{equation}
BW_{f_0(980)} = {1 \over {m_{\pi\pi}^2 - m^2_0 + im_0(\Gamma_{\pi}+\Gamma_K)}}\ ,
\end{equation}
with 
\begin{equation}
\Gamma_{\pi} = g_{\pi}\sqrt{m_{\pi\pi}^2/ 4 - m_{\pi}^2}
\end{equation}
and
\begin{equation}
\Gamma_K = {g_K \over 2}\ \left( \sqrt{m_{\pi\pi}^2/ 4 - m_{K^+}^2}+
\sqrt{m_{\pi\pi}^2/ 4 - m_{K^0}^2}\right)\ .
\end{equation}

We multiply each amplitude  by a complex coefficient, $c_j=a_je^{\delta_j}$.
The fit parameters are the magnitudes, $a_j$, and the phases,
$\delta_j$, which accomodate the final state interactions, are
fit parameters  obtained using the maximum-likelihood method.

Monte Carlo simulations are used to determine the shape and location of the
$D^0 \to K^- \pi^+$  background in the
Dalitz plot. The amount the $D^0$ background is determined using both MC
 simulations and data. The contribution of the  $D_s^+ \to \eta' \pi^+$
 background is negligible. The background proportions are 12 $\pm$ 2\% for 
$D^0 \to K^- \pi^+$ and 88 $\pm$ 2\% for the combinatoric
across the Dalitz plot. Checks of the background model are
described in the companion paper\cite{e791dp}.

The parameters of the $f_0(980)$ state, $g_{\pi}$, $g_K$, and $m_0$, as well 
as the mass and width of the $f_0(1370)$, are determined directly from the 
data, floating them as free parameters in the fit. The other resonance 
masses and widths are taken from the PDG\cite{pdg}. 
The results of the $D_s^+$ Dalitz plot fit are shown in Table I. The column 
labeled Fit A corresponds to our best fit  with all six modes.  
The measured values $m_0 =  977 \pm 3 \pm 2$ MeV/c$^2$, $g_{\pi} =$ 0.09  $\pm$  0.01  $\pm$  0.01 
and  $g_K =$ 0.02  $\pm$  0.04  $\pm$  0.03 have been corrected for small
shifts in these parameters due to the $\pi^-\pi^+$ mass
resolution in this region. This resolution is 
estimated to be approximately 9 MeV/$c^2$. We determine 
the shifts by folding a 9 MeV/$c^2$ resolution 
together with various $g_K$, $g_{\pi}$ combinations and 
fitting the resulting distributions to the form 
in Eqns. (4)-(6).
%
%
The resolution was found to shift $m_0$, $g_K$, and $g_{\pi}$
by $-1.4 $ MeV, +0.06 and $-0.006$ respectively.  These shifts have been included in
the values quoted above and in Table I.  Uncertainties in our resolution contribute
to the systematic uncertainties. The magnitudes and 
phases of the resonant amplitudes are relatively insensitive to the value of $ 
g_K $.
These values are not compatible with WA76 results\cite{wa76},
$g_{\pi} =$ 0.28  $\pm$  0.04 and $g_K =$ 0.56  $\pm$  0.18.
For the $f_0(1370)$ we find $m_0 = $$1434 \pm 18$ MeV/c$^2$
and $\Gamma_0 =  173 \pm 32$ MeV/c$^2$. 
We have also fit the Dalitz plot using for the $f_0(980)$ the same
Breit-Wigner function as for the other resonances. The resulting
fit is nearly as good as the one using the coupled-channel Breit-Wigner function, and the
fractions and phases are indistinguishable. With this parameterization we find 
 $m_0 = $$975 \pm 3$ MeV/c$^2$ and $\Gamma_0 =  44 \pm 2 \pm 2$ MeV/c$^2$, also
 shifted by the effect of mass resolution.

 Table I shows  the magnitudes ($a_j$) and phases
($\delta_j$) determined from the fit, and corresponding fraction for 
each decay mode.
The fractions are defined as

\begin{equation}
f_j \equiv {\int ds_{12} ds_{13} \mid c_j {\cal A}_j \mid^2 \over {\int
ds_{12}ds_{13}
\sum_{jk}  \mid c_j {\cal A}_j c_k^* {\cal A}_k^* \mid}}\ .
\end{equation}

\begin{figure}
\centerline{\epsfysize=3.00in \epsffile{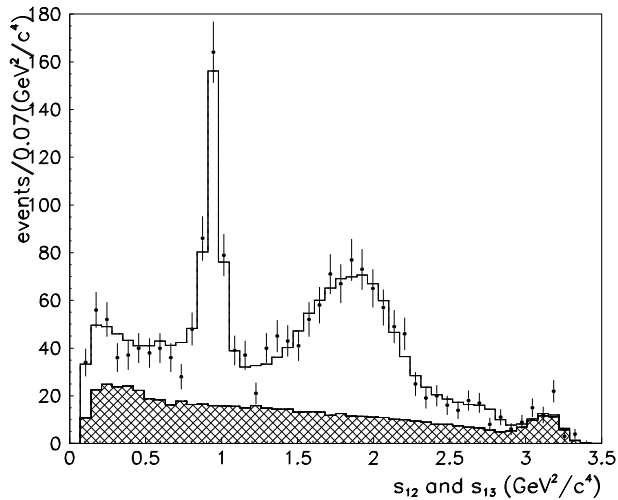}}
\caption{ $s_{12}$ and $s_{13}$ projections for data (dots) and Fit A (solid).
The hashed area is the background distribution.}
\label{fig3}
\end{figure}

\noindent  The first 
reported error is statistical and the second is systematic, the latter being
dominated by the uncertainties in the resonance parameters, in the background
parameterization, and in the acceptance correction. 
The $f_0(980)\pi^+$ is the dominant component, accounting for nearly 
half of the $D_s^+ \to \pi^- \pi^+ \pi^+$ decay width, followed by
the $f_0(1370)\pi^+$ and $f_2(1270)\pi^+$ components. 
The contribution of  $\rho^0(770)\pi^+$ and $\rho^0(1450)\pi^+$ components 
corresponds to about 10\% of the $\pi^- \pi^+ \pi^+$ width. 
We have not found a statistically significant non-resonant component. 
The $s_{12}$ and  $s_{13}$ projections are nearly independent 
and the sum of the two is shown in Fig. \ref{fig3} for Fit A. 
\begin{table}
\centering
 \begin{tabular}{cccc}
\\ \hline     
            &Fit A      &   Fit B     & Fit C \\
	    &Fraction(\%)    &Fraction(\%)    &Fraction(\%)\\
            &Magnitude    &Magnitude    &Magnitude\\
	    &Phase    &Phase    &Phase
 \\ \hline 
\\
$f_0(980)\pi^+$      &56.5  $\pm$  4.3  $\pm$  4.7     &58.0  $\pm$  4.9   &54.1  $\pm$ 
4.0\\ 
                     & 1(fixed)     & 1(fixed)     & 1(fixed) \\
                     & 0(fixed)      & 0(fixed)     & 0(fixed)\vspace{.15cm}\\
NR           &0.5  $\pm$  1.4  $\pm$  1.7               &7.5  $\pm$  4.8     &5.0
 $\pm$  3.8\\
                       &0.09  $\pm$  0.14 $\pm$  0.04    &0.36  $\pm$  0.12    &0.30  $\pm$  0.12  \\ 
                       &(181  $\pm$  94  $\pm$  51)$^{\circ}$    &(165  $\pm$  23)$^{\circ}$   
		       &(149  $\pm$ 25)$^{\circ}$ \vspace{.15cm} \\ 
$\rho^0(770)\pi^+$     &5.8  $\pm$  2.3  $\pm$  3.7         &0    &11.1  $\pm$  2.5\\
                       &0.32  $\pm$  0.07  $\pm$  0.19      &0    & 0.45  $\pm$  0.06 \\ 
                       &(109  $\pm$  24  $\pm$  5)$^{\circ}$       &0    &(81  $\pm$  15)$^{\circ}$ \vspace{.15cm} \\ 
$f_2(1270)\pi^+$       &19.7  $\pm$  3.3  $\pm$  0.6    &22.2  $\pm$  3.3    &20.8  $\pm$  3.0\\
                       &0.59  $\pm$  0.06  $\pm$  0.02    &0.62  $\pm$  0.06     & 0.62  $\pm$  0.05 \\
                       &(133  $\pm$  13  $\pm$  28)$^{\circ}$     &(109  $\pm$  11)$^{\circ}$ 
    &(124  $\pm$  11)$^{\circ}$ \vspace{.15cm}\\
$f_0(1370)\pi^+$       &32.4  $\pm$  7.7  $\pm$  1.9    &30.4  $\pm$  6.9   &34.7
 $\pm$  7.2\\
                       &0.76  $\pm$  0.11  $\pm$  0.03     &0.72  $\pm$  0.11    &0.80  $\pm$ 0.11 \\
                       &(198  $\pm$  19  $\pm$  27)$^{\circ}$    &(156  $\pm$  19)$^{\circ}$   
		       &(159  $\pm$  14)$^{\circ}$\vspace{.15cm}\\ 
$\rho^0(1450)\pi^+$    &4.4  $\pm$  2.1  $\pm$  0.2    &5.8  $\pm$  2.2   &0\\
                       &0.28  $\pm$  0.07  $\pm$  0.01    &0.32  $\pm$  0.06    & 0 \\
                       &(162  $\pm$  26  $\pm$  17)$^{\circ}$    &(144  $\pm$  20)$^{\circ}$    &0\vspace{.15cm}\\
$m_{f_0(980)} $         & 977 $\pm$ 3 $\pm$  2   &  976  $\pm$  3 & 974 $\pm$ 3\\
{\footnotesize (MeV/c$^2$)}& & & \\
$g_\pi$                 &0.09 $\pm$ 0.01 $\pm$ 0.01     & 0.10  $\pm$  0.01  & 0.09 $\pm$ 0.01\\

$g_K$                  &0.02 $\pm$ 0.04  $\pm$ 0.03     & -0.02 $\pm$  0.04 & 0.04 $\pm$ 0.04\\
$m_{f_0(1370)}$       &1434 $\pm$ 18 $\pm$ 9   & 1401 $\pm$  19   &
1406 $\pm$ 15\\
{\footnotesize (MeV/c$^2$)}& & & \\
$\Gamma_{f_0(1370)}$  &172 $\pm$ 32 $\pm$ 6     &180  $\pm$  34   & 176 $\pm$ 30\\
{\footnotesize (MeV/c$^2$)}& & & \\

\\$\chi^2/ \nu$     & 71.8/68      & 93.0/68     & 103.5/68 
\\ C.L.     & 35\%     &2\%     & 0.4\% 
\\ -2 ln${\cal L}_{max}$    &-3204   &-3184    & -3172
\\ \hline
\\
\end{tabular}
\protect\caption{Results of the $D_s$ Dalitz plot fits. Fit A,  all resonances are allowed
(systematic error follows the statistical). Fit B and Fit C do not include the $\rho^0(770)\pi^+$
and $\rho^0(1450)\pi^+$ amplitudes, respectively.}
\label{tab1}
\end{table}

To assess the quality of our fit absolutely, and to compare it with other 
possible fits, we developed a fast-MC algorithm that simulates the 
$D_s^+ \to \pi^- \pi^+ \pi^+$ Dalitz plot from a given signal distribution,
background, detector resolution and acceptance.
For any given set of input parameters we calculated a $ \chi^2 $ using the
procedure presented in Ref. \cite{e791dp}.  
From $ \chi^2 $  and the number of degrees of freedom ($ \nu $),
we calculate a confidence level assuming a Gaussian distribution
in $ \chi^2 / \nu $.
The confidence level for the agreement of the projection of Fit A onto 
the Dalitz plot with the
data  of Fit A is 35\%.

We perform fits excluding amplitudes with small contributions. The fit without
the non-resonant amplitude is as good as Fit A, and the resulting parameters 
are essentially the same. When comparing Fit A with models without the
$\rho^0\pi^+$ amplitudes (Table 1) we  calculate 
$ \Delta w = -2 ({\rm ln} {\cal L}_i - {\rm ln} {\cal L}_A)$, where
${\cal L}_A$ is the likelihood of the fit with all modes, and
${\cal L}_i$ is the likelihood of the different models, as described in the
companion paper\cite{e791dp}.
For the model where we exclude the $\rho^0(770)\pi^+$  amplitude we have, for
data, $\Delta w = $20. In the fast-MC with all modes we have
$\langle \Delta w \rangle  = 37$. In the fast-MC with no 
$\rho^0(770)\pi^+$ we have $\langle \Delta w \rangle  = -29$.
We observe a similar
behavior in Fit C, where we exclude the $\rho^0(1450)\pi^+$  amplitude,
we have  $\Delta w = 32$. In the fast-MC with all modes we have 
$\langle \Delta w \rangle  = 30$ while with 
 the fast-MC with no $\rho^0(1450)\pi^+$ 
$\langle \Delta w \rangle  = -30$.
In all fast-MC experiments the rms deviation for 
$\langle \Delta w \rangle$ is about 11 units. 
We conclude that the best description of our data includes both  
$\rho^0(770)\pi^+$ and $\rho^0(1450)\pi^+$ amplitudes. 

The contribution of modes having isoscalar mesons completely 
dominates the $D_s^+ \to \pi^- \pi^+ \pi^+$ decay. The same 
isoscalar dominance is  observed in the $D^+ \to \pi^- \pi^+ \pi^+$ 
decay\cite{e791dp}.  
However, there is no evidence in the $D_s^+$ decay for a low-mass
broad scalar particle as seen in the  $D^+$ decay.
If the $D_s^+ \to \pi^- \pi^+ \pi^+$
decay is dominated by the Cabibbo-favored spectator mechanism, we would
expect final states with a large $s\bar{s}$  content.  
Approximately half of the $D_s^+ \to \pi^- \pi^+ \pi^+$ 
rate is produced via  $f_0(980)\pi^+$. The $ f_0(980) $ is often 
supposed to have a large $ s \overline s $ component, indicating 
a large spectator amplitude in this decay. On the other hand, the 
 large contribution from the  intermediate state  $ f_0 (1370) \pi^+$ 
 indicates the  presence of either $ W$-annihilation amplitudes or strong 
 rescattering in the final state. In fact this decay  is not observed
 in the $D_s^+ \to K^+K^-\pi^+$  final state\cite{e687kkpi}, pointing to the
 $f_0(1370)$ being  a  
 non-$ s \overline s $  particle, as suggested by the 
 naive quark model\cite{pdg}.

In summary, Fermilab experiment E791 has measured the branching ratio of
the decay  $D_s^+ \to \pi^- \pi^+ \pi^+$ relative to $D_s^+ \to \phi \pi^+$
to be $0.245 \pm 0.028^{+0.019}_{-0.012}$. We  measure the mass and
width of the $f_0(980)$ and $f_0(1370)$. Our results for
the $f_0(980)$ parameters are $g_{\pi} =$ 0.09   $\pm$  0.01  $\pm$  0.01, 
$g_K =$0.02  $\pm$  0.04  $\pm$  0.03, and $m_0 = 977 \pm 3 \pm 2 $ MeV/c$^2$
Using the same Breit-Wigner function for the $f_0(980)$ as for the other 
resonances, we find 
$\Gamma_0 =  44 \pm 2 \pm 2$ MeV/c$^2$ and $m_0=975\pm 3$~MeV/c$^2$. For the $f_0(1370)$ we find 
$m_0 = $$1434 \pm 18 \pm 9$ MeV/c$^2$
and $\Gamma_0 =  173 \pm 32 \pm 6$ MeV/c$^2$. Finally, the fit of the Dalitz plot
shows a dominant contribution from the $f_0(980)\pi^+$, significant
contributions from the $f_0(1370)\pi^+$ and $f_2(1270)\pi^+$,
small contribution from the $\rho^0 \pi^+$ channels, and a negligible 
non-resonant component. The isoscalar plus $\pi^+$ components correspond to
over 90\% of the $D_s^+ \to \pi^- \pi^+ \pi^+$ decay width.

We gratefully acknowledge the assistance of the staffs of Fermilab and of all
the participating institutions.  This research was supported by the Brazilian
Conselho Nacional de Desenvolvimento Cient\'{\i}fico e Tecnol\'{o}gico,
CONACyT (Mexico), the Israeli Academy of Sciences and Humanities, 
the U.S. Department of Energy, the U.S.-Israel
Binational Science Foundation, and the U.S. National Science Foundation.
Fermilab is operated by the Universities Research Association, Inc., under
contract with the United States Department of Energy.

\small
\bibliographystyle{unsrt}

\begin{thebibliography}{99.}
\small


\bibitem{pdg} 
Particle Data Group, C. Caso {\em{et al.}}, 
Eur. Phys. J. C {\bf 3}, 1 (1998).
 

\bibitem{e691}
E691 Collaboration, J.C.~Anjos {\em{et al.}}, Phys. Rev. Lett. {\bf 62}, 125 (1989).

\bibitem{e687} E687 Collaboration, P.L.~Frabetti, {\em{et al.}} Phys. Lett. B 
{\bf407},79 (1997).

\bibitem{mont} L. Montanet, Nucl. Phys. B {\bf 86}, 381 (2000).

\bibitem{e791dp} E791 Collaboration, E.M.~Aitala {\em{et al.}}, ``Experimental
 evidence of a light and broad scalar resonance in 
$D^+\to \pi^- \pi^+ \pi^+$ Decay''(Following  paper).

\bibitem{OLDTPL}E791 Collaboration, E.M.~Aitala {\em{et al.}}, Eur.Phys.J 
Direct C{\bf  4}, 1 (1999).

\bibitem{wa82} WA82 Collaboration, M.~Adamovich  {\em{et al.}}, Phys. Lett. B
{\bf 305},  177 (1993).



\bibitem{blatt} J.M.~Blatt and V.F.~Weisskopf, Theoretical Nuclear Physics,
John Wiley \& Sons, New York, 1952.

\bibitem{wa76} 
WA76 Collaboration, T.A.~Armstrong {\it et al.},
Z. Phys. {\bf C  51}, 351 (1991); E.~Levin, CBPF-FN-009/95 (unpublished).

\bibitem{e687kkpi} E687 Collaboration, P.L.~Frabetti  {\it et al.},
Phys. Lett. B {\bf 351}, 591 (1995). 




\end{thebibliography}

\end{document}